\documentclass[aps,prl,twocolumn,superscriptaddress,nofootinbib]{revtex4-2}

\usepackage{amsmath,amssymb,amsfonts}
\usepackage{mathtools}
\usepackage{graphicx}
\usepackage{hyperref}
\usepackage{bm}
\usepackage{xcolor}

\newcommand{\Gr}{\mathrm{Gr}}
\newcommand{\PP}{\mathbb{P}}

\newcommand{\cI}{\mathcal{I}}

\begin{document}

\title{Positivity and Cluster Structures in Landau Analysis}

\author{Benjamin Hollering}
\affiliation{Max Planck Institute for Mathematics in the Sciences, Leipzig, Germany}

\author{Elia Mazzucchelli}
\affiliation{Max Planck Institute for Physics, Garching, Germany}

\author{Matteo Parisi}
\affiliation{Okinawa Institute of Science and Technology, Okinawa, Japan}
\affiliation{Max Planck Institute for Physics, Garching, Germany}

\author{Bernd Sturmfels}
\affiliation{Max Planck Institute for Mathematics in the Sciences, Leipzig, Germany}

\begin{abstract}

Landau analysis in momentum twistor space can be formulated as the study of varieties of lines in three-dimensional projective space, together with their projections and discriminants. Within this framework, we define enumerative invariants (LS degrees) that count leading singularities. Leading Landau singularities (LS discriminants) arise as discriminants detecting the collision of leading singularities.
We uncover a recursive mechanism underlying Landau singularities, governed by substitution maps between Grassmannians. Applying this framework, we prove positivity and factorization into cluster variables for the LS discriminant of a large class of Landau diagrams at arbitrary loop order. This provides a first-principles explanation for the emergence of positivity and cluster algebra structures in the singularities of planar $\mathcal{N}=4$ super Yang–Mills theory.

\end{abstract}

\makebox[0pt][l]{\hspace*{6.5cm}\raisebox{0.9cm}[0pt][0pt]{\small MPP-2026-48}}

\maketitle

\section{Introduction}

Modern progress in perturbative quantum field theory relies increasingly on computing scattering amplitudes and understanding their singularity structure.
Scattering amplitudes are built from integrals of the form \begin{equation}
\cI(\mathbf{M})\;=\;\int_{\Gamma}
\frac{N(\mathbf{L};\mathbf{M})}{D(\mathbf{L};\mathbf{M})}\, d\mu(\mathbf{L}) \, .
\label{eq:integral}
\end{equation}
The ingredients are the
 loop variables $\mathbf{L}$, external kinematics  $\mathbf{M}$,
integration measure $\mu$, and contour $\Gamma$. The explicit form of~\eqref{eq:integral} depends on the choice of kinematic variables~\cite{Weinzierl2022}.
In the case of a Feynman integral in momentum representation, $D(\mathbf{L};\mathbf{M})$ is the product of propagators associated with the internal edges of the corresponding Feynman diagram.
The integral $\mathcal{I}(\mathbf{M})$ is a transcendental function involving e.g.~multiple polylogarithms, elliptic integrals, Calabi--Yau periods.
A first step towards computing $\mathcal{I}(\mathbf{M})$ is determining its singularities, namely the locus in the external kinematics where $\mathcal{I}(\mathbf{M})$ develops poles or branch cuts.  
This locus is the \textit{Landau variety}.

The core idea of \textit{Landau analysis}~\cite{Landau1960,Hannesdottir:2025bootstrap,helmer2024landau,Eden:1966dn,Mizera:2022landau,Hannesdottir:2022iterated,Fevola:2024principal,Fevola:2024revisited,Caron-Huot:2025recursive,He:2024schubert,Correia:2025sofia,Dennen:2017}
is the determination of the Landau variety from the singular locus $\{D(\mathbf{L};\mathbf{M})=0\}$ of the integrand in \eqref{eq:integral}, and its interplay with the cycle $\Gamma$~\cite{Pham2011}. 
In this letter we use \textit{momentum twistor} variables~\cite{hodges2013eliminating} to exploit the rich algebro-geometric structure of the Grassmannian. In these variables, vanishing propagators give incidence conditions for pairs of lines in projective $3$-space. Hence on-shell spaces are subvarieties in a product of Grassmannians of lines~\cite{Our_lines_1,Our_Landa_on_Gr}. 

The Landau variety is the branch locus of the \textit{Landau map}, the projection from the on-shell space to the space of external kinematics~\cite{Landau1960}. When the Landau map has finite
fibers, we call its branch locus~\textit{leading Landau singularity}.  
The points in the fiber correspond to maximal residues of the integrand in~\eqref{eq:integral}.  
These \textit{leading singularities}  appear as maximal iterated discontinuities of~\eqref{eq:integral} and as algebraic prefactors in analytic evaluations~\cite{Cachazo:2008vp}. 

Momentum twistor variables are well-suited for describing planar dual conformal invariant integrals \cite{Drummond:2010qh}.
Our results have important implications for scattering amplitudes in planar $\mathcal{N}=4$ super Yang-Mills (SYM)~theory.  
Here additional geometric structures arise, such as positive geometries and cluster algebras~\cite{The_amplituhedron,Golden:2013xva}.  
At fixed multiplicity and loop order, the full color-ordered amplitude can be written as a single integral \eqref{eq:integral}, 
whose~integrand is the \textit{canonical form} of the loop amplituhedron~\cite{The_amplituhedron}.

The connection between scattering amplitudes and cluster algebras was first observed by Golden \emph{et al.}~\cite{Golden:2013xva}.
%
This was followed by the discovery of \emph{cluster adjacencies}~\cite{DFG} and related patterns~\cite{GP23}, culminating in the cluster bootstrap program~\cite{CHDDDFGvHMP}. Despite these advances, a first-principles explanation for the emergence of cluster structures in particle physics has remained elusive.
At tree level, where amplitudes are rational functions, recent progress~\cite{EZLPTSW} has explained the appearance of cluster structures in terms of the positive geometry of the amplituhedron. A key ingredient is the introduction of \emph{BCFW promotion maps}.
These \emph{cluster quasi-homomorphisms} mirror the BCFW recursive construction of amplitudes.
 
\textbf{Main Contributions.}
We develop a framework for \textit{Landau analysis on the Grassmannian}, in which (leading and super-leading) Landau singularities are expressed as discriminants and resultants. We uncover recursive structures governing these objects, whose key ingredient is a class of \textit{substitution maps} closely related to \textit{promotion maps} for cluster algebras~\cite{PlabicTangles}. This provides a pathway toward proving two central conjectures about singularities of planar $\mathcal{N}=4$ SYM amplitudes: \emph{positivity}, namely regularity in the positive kinematic region, and the emergence of cluster algebras, which we formulate as \emph{cluster factorization} of Landau singularities into cluster variables in the rational regime. The broader geometric and algebraic constructions are developed in~\cite{Our_Landa_on_Gr}.

We prove both positivity and cluster factorization at all loop orders for planar Landau diagrams whose \textit{dual loop graph} $G$ is a tree, i.e. the subgraph of the dual diagram whose vertices correspond to loop variables.
In addition, we perform extensive numerical checks for planar Landau diagrams up to four loops. Our approach yields a first-principles explanation for the emergence of positivity and cluster structures in planar $\mathcal{N}=4$ SYM amplitudes, while also providing effective tools for the computation of Landau singularities.


\section{On-Shell Spaces and the Landau map}

The Grassmannian $\Gr(2,4)$ is the space of lines in 
projective $3$-space $\PP^3$. In momentum twistors, propagators become incidences of lines~\cite[Section 2]{Our_Landa_on_Gr}, and on-shell spaces~\cite{Sequential_Discontinuitites} are  subvarieties in  $\Gr(2,4)^\ell$.
Fix $\ell \geq 1$ and consider a connected undirected simple graph $G$ on $\ell$ vertices. 
We define the \textit{line incidence variety $V_G \subseteq \Gr(2,4)^\ell$ of }$G$ as the configurations of $\ell$ lines $\mathbf{L}=(L_1, \dots,L_\ell)$ satisfying the incidence $\langle L_iL_j\rangle =0$ for every edge $ij$ of $G$.

Questions about the irreducibility, dimension and complete intersection of $V_G$
were resolved in~\cite[Sect 5]{Our_lines_1}.
Both answers rely on the rigidity theory of graphs in the plane \cite{SS}. For example, the variety of the triangle $V_{K_3}$ is
$9$-dimensional and has two irreducible components $V_{[3]}$ and $V_{[3]}^*$, of concurrent and coplanar lines. This geometric understanding is crucial for Landau analysis as in~\cite{helmer2024landau}.

To the $\ell$ \textit{loop} lines $\mathbf{L}$ we adjoin $d$~\emph{external} lines $\mathbf{M}=(M_1,\dots,M_d)$.
Let $u\in \mathbb{N}^\ell$ with $|u|=\sum_{i=1}^\ell u_i = d$ and form a graph $G_u$ by attaching $u_i$ pendant edges to vertex $i$.
Each pendant edge corresponds to an external line incident to $L_i$.  We call $\mathcal{L}=G_u$ a \textit{Landau diagram};
this is actually the planar dual of a usual Landau diagram~\cite{Eden:1966dn}.
The line incidence variety $V_{\mathcal{L}} \subseteq \Gr(2,4)^{\ell+d}$ is the on-shell space of $\mathcal{L}$: it contains the lines (kinematics) satisfying all incidences (cut or on-shell conditions) prescribed by~$\mathcal{L}$.

Given a Landau diagram $\mathcal{L}=G_u$, consider the map
\begin{equation}
\psi_\mathcal{L} :\ V_{\mathcal{L}}\rightarrow \Gr(2,4)^d \, ,
\quad
(\mathbf{L};\mathbf{M})\mapsto \mathbf{M} \, .
\label{eq:Psi}
\end{equation}
The key idea from Landau analysis is that the Landau variety is recovered as the union of the branch loci of $\psi_\mathcal{L}$, where  $G$ and $u$ vary within a family determined by the structure of~\eqref{eq:integral}~\cite{Pham2011}. For fixed external lines $\mathbf{M}$, the fiber $\psi_\mathcal{L}^{-1}(\mathbf{M})$ consists of all loop lines satisfying the on-shell conditions associated to edges of $\mathcal{L}$. The Landau variety detects the locus in the external kinematics where the fiber is non-generic~\cite{Pham2011}. We focus on \textit{leading Landau diagrams} $\mathcal{L}$, when the generic fiber of $\psi_{\mathcal{L}}$ consists of finitely many points. 
We address the following questions:
\begin{enumerate}
\item \emph{Size:} What is the cardinality of the generic fiber?
\vspace{-0.09in}
\item \emph{Discriminants:} Which $\mathbf{M}$ have non-generic fiber?
\vspace{-0.09in}
\item \emph{Reality:} For which real $\mathbf{M}$ is the whole fiber real?
\vspace{-0.09in}
\item \emph{Rationality:} For which constraints on the external data 
are all points in the fiber rational in~$\mathbf{M}$?
\end{enumerate}

\section{Leading singularity degree}

Fix a graph $G$ and  $u \in \mathbb{N}^\ell$  with
$|u|=d=\dim(V_G)$.
We form a Landau diagram $\mathcal{L}=G_u$ with Landau map $\psi_u=\psi_{\mathcal{L}}$. 
 We expect the fiber $\psi_u^{-1}(\mathbf{M})$ to be empty or
 finite. Its points are configurations of $\ell$ lines, fully localized by imposing the on-shell conditions.
 The points are the \textit{leading singularities}. We define the \emph{LS degree} by
\begin{equation}
\gamma_u\;:=\;\big|\psi_u^{-1}(\mathbf{M})\big|
\quad \text{for generic } \mathbf{M} \, .
\label{eq:LSdegree}
\end{equation}
The LS degrees count solutions to Schubert problems, i.e. incidence problems for lines in projective space. If $G=K_1$ is the one-vertex graph and $u=(4)$, then $\gamma_{(4)}=2$: 
four generic lines in $\mathbb{P}^3$ have two transversals.

From an algebraic viewpoint, $\gamma_u$ is a coefficient in the multidegree $[V_G]$, see~\cite[Sect~3]{Our_lines_1}.
This polynomial represents a cohomology class and packages, for the fixed dual loop graph $G$, all leading Landau diagrams $G_u$ and their LS degrees $\gamma_u$.
When $V_G$ is a complete intersection~\cite[Thm 5.4]{Our_lines_1}, meaning $\dim(V_G) = 4 \ell - |G|$, then
\begin{equation*}
    [V_G] = 2^\ell \cdot \prod_{i=1}^\ell t_i \cdot \prod_{ij \in G} (t_i + t_j) = \sum_{u \in \mathbb{N}^\ell} \gamma_u \, t_1^{5-u_1} \cdots t_\ell^{5-u_\ell} \, .
\end{equation*}
For example, 
$\,
   [V_G] = 16 \cdot t_1^2 t_2^2 t_3^2 + 8 \cdot t_1^3 t_2^2 t_3 + {\rm perm.} \,$ for $G=K_3$.
The coefficient $16$ is the LS degree of triple pentagon $G_u$ with $u=(3,3,3)$~\cite{Higher_roots}.
For all graphs up to $8$ vertices, i.e. leading Landau diagrams with up to $8$ loops (planar and non-planar), we determine the  dimension, irreducible decomposition, and multidegree in~\cite{Our_lines_1}. The data is presented at
\hbox{\url{https://zenodo.org/records/17708048}}.


\section{Landau discriminants and resultants}

We focus on \textit{leading} and \textit{super-leading} Landau singularities, where $d-\dim(V_G)  $ equals $0$ or $1$ respectively.
If  $\mathcal{L}$ is a leading Landau diagram $(|u|=d)$ then the branch locus of $\psi_\mathcal{L}$ detects when leading singularities collide.
This property is encoded by the \emph{LS discriminant} $\Delta_\mathcal{L}(\mathbf{M})$:
\begin{equation}
\!\!\!\! \Delta_\mathcal{L}(\mathbf{M})=0
\quad {\rm whenever} \quad
\big|\psi_\mathcal{L}^{-1}(\mathbf{M}) \big|<\text{generic value}.
\label{eq:discriminant}
\end{equation}
This is an analog of the univariate discriminant.
For instance, the LS discriminant of the \textit{four-mass box} with $G{=}K_1$ and $u{=}(4)$ is the $4 {\times} 4$ Gram 
determinant $\langle M_i M_j \rangle$.

Let $\mathcal{L}$ be a super-leading Landau diagram. We have $\psi_{\mathcal{L}}^{-1}(\mathbf{M}) = \emptyset$
for generic $\mathbf{M}$.
The \emph{SLS resultant} satisfies
\begin{equation}
R_\mathcal{L}(\mathbf{M})=0
\quad {\rm whenever}\quad
\psi_\mathcal{L}^{-1}(\mathbf{M}) \neq \emptyset \, .
\label{eq:resultant}
\end{equation}
This captures when an overdetermined polynomial system admits a solution.
As an example, the SLS resultant of the \textit{pentagon} with $G=K_1$ and $u=(5)$ is the determinant of the $5 \times 5$ Gram matrix $\langle M_i M_j \rangle$ as above. 

    Our full paper~\cite{Our_Landa_on_Gr} reports on computations of Landau discriminants and resultants. We
    give insight into leading and super-leading singularities of amplitudes and Feynman integrals. We also study the degrees of discriminants and resultant~\cite[Thm 4.4, Thm 5.1, Prop 5.3]{Our_Landa_on_Gr}.  Lastly, we look into \textit{next-to-leading (NLS)} singularities, where the generic fibers of the Landau map are curves: we give a formula for their genera~\cite[Section 6]{Our_Landa_on_Gr} and compute the Landau singularity of the double box with all external momenta off shell.



\section{Recursive Landau Analysis}

One of our 
key discoveries is that many Landau singularities have a \emph{recursive} structure via graph decompositions. Our approach is 
more algebro-geometric than~\cite{Caron-Huot:2025recursive}.

\begin{figure}[h!]
    \centering
\includegraphics[width=0.47\textwidth]{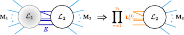}
    \caption{The recursion for Landau singularities of a Landau diagram $\mathcal{L}$ that is reducible by a leading Landau diagram $\mathcal{L}_1$.}
    \label{fig:recursion}
\end{figure}

Consider a Landau diagram which decomposes as
$\mathcal{L} = G_u \cup H_u$, $\mathcal{L}_1 = G_{1,u_1} \cup H_{1,u_1}$ and $\mathcal{L}_2 = G_{2,u_2} \cup H_{2,u_2} \cup E$ such that $u=(u_1,u_2)$, $G=G_1 \cup G_2 \cup E$ with $E$ the edges between $G_1$ and $G_2$. Assume that $\mathcal{L}_1$ is \textit{leading}, which means that $|u_1| = \dim(V_{G_1})$. We say that $\mathcal{L}$ is \textit{reducible by} $\mathcal{L}_1$. Write $\mathbf{M}=(\mathbf{M}^{(1)},\mathbf{M}^{(2)})$ for the $|u_1|+|u_2|$ external lines. Solving the on-shell conditions for $\mathcal{L}_1$ for fixed $\mathbf{M}^{(1)}$ already fully localizes the loop lines in $G_1$ to $\gamma_1$ configurations 
$\mathbf{L}^{(1)}_r=\mathbf{L}^{(1)}_r(\mathbf{M}^{(1)})$, where $\gamma_1$ is the LS degree of $\mathcal{L}_1$
and $r=1,\dots,\gamma_1$.
This structure yields a decomposition of the on-shell space, and hence of the fiber of the Landau map $\psi$. As a result, the branch locus of $\psi$ can be characterized in terms of that of the Landau maps $\psi_i$ on $\mathcal{L}_{i}$. A branch point $\mathbf{M}=(\mathbf{M}^{(1)},\mathbf{M}^{(2)})$ of $\psi$  is such that: either $\mathbf{M}^{(1)}$ is a branch point of $\psi_1$, or $(\mathbf{L}^{(1)}_r,\mathbf{M}^{(2)})$ is a branch point for $\psi_2$ for some $r$. 
This also implies an algebraic factorization: if ${\rm LS}_{\mathcal{L}}$ and ${\rm LS}_{\mathcal{L}_i}$ are the Landau singularities of $\mathcal{L}$ and $\mathcal{L}_i$, then
\begin{equation}\label{eq:recursion}
    {\rm LS}_{\mathcal{L}}(\mathbf{M}) = \Delta_{\mathcal{L}_1}(\mathbf{M}^{(1)}) \cdot \mathcal{E}(\mathbf{M}^{(1)}) \cdot   \prod_{r=1}^{\gamma_1} \varphi_{r}^* \, {\rm LS}_{\mathcal{L}_2}(\mathbf{M}) \, .
\end{equation}
Each \textit{substitution map} $\varphi_{r}^* $ inserts the lines $\mathbf{L}^{(1)}_r(\mathbf{M}^{(1)})$ in the polynomial ${\rm LS}_{\mathcal{L}_2}$ according to $\mathcal{L}$, see Fig.~\ref{fig:recursion}. The \textit{extraneous factor} $\mathcal{E}(\mathbf{M}^{(1)})$ is a rational function in $\mathbf{M}^{(1)}$ reflecting the (projective) ambiguity in how we define the substitution maps; by possibly redefining $\varphi_r$, we can set $\mathcal{E}(\mathbf{M}^{(1)})=1$. Despite each factor $\varphi_{r}^* \, {\rm LS}_{\mathcal{L}_2}$ in~\eqref{eq:recursion} 
being algebraic, the full product is rational~\cite[Thm 7.4]{Our_Landa_on_Gr}.

We now focus on \textit{leading} and \textit{super-leading} Landau diagrams $\mathcal{L}$. Here, ${\rm LS}_\mathcal{L}$ is replaced by $\Delta_\mathcal{L}$ and $R_\mathcal{L}$, respectively.  
This recursion is the main engine behind our positivity and cluster structure results,  discussed below. 

For the four-mass box $\mathcal{L}_1$ with $G_1=K_1$ and $u_1=(4)$,
we have $\gamma_1=2$ points $\mathbf{L}^{(1)}_{\pm}=(L_{\pm})$
in the Landau fiber over
$\mathbf{M}^{(1)}=(A,B,C,D)$; see \cite[Eq. (26)]{Our_Landa_on_Gr}.
Namely, $L_{\pm}  = \varepsilon \cdot \left(p(x^{(\pm)}) B \right)  \star \left(p(x^{(\pm)})C \right)  $ where $p(x) = d_1 + x d_2$ parametrizes the line $D=d_1d_2$ and $\star$ is the geometric intersection. The roots $x^{(\pm)}  = (-a_1 \pm \sqrt{ a_1^2 - 4 a_0 a_2})/(2 a_2)$ are written in terms of \textit{chain polynomials} $\langle abc | de | fgh \rangle:= \langle abcd \rangle \langle efgh \rangle- \langle abce \rangle \langle dfgh \rangle$. Namely, $a_2 =  \langle d_2 A | B | d_2 C \rangle$, $a_1 =  \langle d_1 A | B | d_2 C \rangle +  \langle d_2 A | B | d_1 C \rangle$ and $ a_0 =  \langle d_1 A | B | d_1 C \rangle$. The normalization $\varepsilon = 2 a_2/\sqrt{\langle B C \rangle \langle B D \rangle \langle CD \rangle}$ is chosen such that $L_{\pm}$ have degree $(1,1,1,1)$ in $(A,B,C,D)$. 

\begin{figure}[h!]
    \centering
\includegraphics[width=0.45\textwidth]{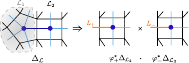}
    \caption{Recursion for the LS discriminant  of the pentabox.}
    \label{fig:pentabox}
\end{figure}

\paragraph{Example ($\ell=2$).}
Let $\mathcal{L}=G_u$ be the pentabox with $G=K_2$ and $u=(4,3)$. This is reducible by the four-mass box $\mathcal{L}_1$ and~\eqref{eq:recursion} yields (omitting the pre-factor $\Delta_{\mathcal{L}_1}$):
\begin{equation}\label{eq:recursion_pentabox_LS}
    \Delta_{\mathcal{L}}(\mathbf{M}) =  \prod_{\sigma \in \{\pm\}} \Delta_{K_1,(4)}(\varphi_{\pm}(\mathbf{M})) \, ,
\end{equation}
The \textit{four-mass box substitution maps} $\varphi_{\pm}$ are defined by $\varphi_{\pm}(\mathbf{M}) = (L_{\pm}(\mathbf{M}^{(1)}),\mathbf{M}^{(2)})$. This is depicted in Fig.~\ref{fig:pentabox}.

The same recursion holds for the SLS resultant (indicating super-leading Landau singularities). Here $\mathcal{L}$ is the double box, $G=K_2$ and $u=(4,4)$. Then, \eqref{eq:recursion_pentabox_LS} holds by replacing $\Delta_{K_2,(4,3)}$ by $R_{K_2,(4,4)}$, and $\Delta_{K_1,(4)}$ by $R_{K_1,(5)}$.


In general, if
$\mathcal{L}$ is reducible by the four-mass box $\mathcal{L}_1$, 
\begin{equation}\label{eq:four_mass_recursion}
    \Delta_{\mathcal{L}}(\mathbf{M}) \, \,  = \, \,  \prod_{\sigma \in \{\pm \}} \Delta_{\mathcal{L}_2}(\varphi_{\pm}(\mathbf{M})) \, .
\end{equation}
A similar formula holds for SLS resultants if $\mathcal{L}$ is super-leading. 
This works well for all trees $G$. We stress that the absence of the extraneous factor crucially depends on our choice of $L_{\pm}$, and hence of $\varphi_{\pm}$, as in~\cite[Eq. (26)]{Our_Landa_on_Gr}. 


\section{Reality and Positivity}

We consider a notion of positivity on the external data.
Assume that $\mathcal{L}$ is planar and embedded in a disk, so that
the $d$ external lines $\mathbf{M}$ have a circular order.
Write $M_i = z_{p_i} z_{p_{i+1}}$ and collect the points $z_p$ in a $4 \times n$ matrix $\mathbf{z}=[z_1 | \cdots |z_n]$ with $n \leq 2d$.
We remove equal columns if some external lines intersect. We
call $\mathbf{M}$ is \textit{positive} if $\mathbf{z}\in {\rm Gr}_{>0}(4,n)$, the positive Grassmannian, i.e.\ the subset of $\Gr(4,n)$ represented by matrices with all ordered maximal minors positive. This notion of positivity is expected to yield a vast family of real Schubert problems\vspace{1.5mm}.  

\textbf{Reality Conjecture:}
If $\mathcal{L}$ is planar leading and $\mathbf{M}$ is positive, then all points in the fiber $\psi_\mathcal{L}^{-1}(\mathbf{M})$ are real\vspace{1.5mm}.

Reality is tightly related to positivity. A rational function on the  Grassmannian ${\rm Gr}(k,n)$ is called \textit{Grassmann copositive} if it is positive on ${\rm Gr}_{>0}(k,n)$. Non-trivial examples
are the LS discriminant $\Delta_{K_1,(4)}$ of the box and the SLS resultant of the pentagon $R_{K_1,(5)}$, see~\cite[Prop 9.2]{Our_Landa_on_Gr}. All amplitudes in planar $\mathcal{N}=4$ SYM are expected to be regular on the positive Grassmannian ${\rm Gr}_{>0}(4,n)$.

\vspace{1.2mm}

\textbf{Positivity Conjecture:}
If $\mathcal{L}$ is planar leading (or super-leading) then $\Delta_{\mathcal{L}}$ (or $R_{\mathcal{L}}$) is Grassmann copositive\vspace{1.0mm}.

\smallskip
Our result in~\cite[Thm 9.4, Cor 9.7, Prop 9.8.]{Our_Landa_on_Gr} states
\vspace{1.0mm}:

\textbf{Theorem} (Reality and Positivity for Trees) Both the Reality and Positivity Conjectures are true if $G$ is a tree\vspace{1.0mm}.

 The key to the proof is that the substitution maps $\varphi_{\pm}$ in~\eqref{eq:recursion_pentabox_LS} and \eqref{eq:four_mass_recursion} are \textit{copositive},
 i.e.~they map positive lines  $\mathbf{M}$ to positive lines \cite[Lemma 9.3]{Our_Landa_on_Gr}. 
 Together with the recursive structure seen above, one 
 derives both conjectures inductively 
 on the number of vertices of~$G$.

The Positivity Conjecture for LS discriminants and resultants are
meant for each irreducible component of~$V_G$ separately; see \cite[Section 10]{Our_Landa_on_Gr} for  details. 
We verified the Reality Conjecture 
computationally for all planar graphs $G$ on $\ell =3, 4$ vertices with {\tt HomotopyContinuation.jl}  \cite{BT}. For each $G$ and $u$ such that $\gamma_u \neq 0$, we checked for $10^5$ random $\mathbf{M}$ that $\psi_{\mathcal{L}}^{-1}(\mathbf{M})$ has only real points.
Similarly, we checked copositivity of the substitution maps.
Hence $G$ is a  `seed' for the recursion in~\eqref{eq:recursion} 
that yields Positivity for other infinite families of Landau diagrams.


\section{Rationality and Cluster Structure}

We now degenerate the external data $\mathbf{M}$ so that
all points in the Landau fiber are rational functions of $\mathbf{M}$.
We set some of the external lines to be incident by fixing a graph $H_u$ on $d$ vertices, and restricting ${\bf M}$ to lie in $V_{H_u}\subseteq {\rm Gr}(2,4)^d$. We denote the resulting Landau diagram by $\mathcal{L}=G_u\cup H_u$ and its Landau map by
$\psi_\mathcal{L}:V_{\mathcal{L}}\rightarrow V_{H_u}$. We call $\mathcal{L}$ rational if it is a leading 
Landau diagram and each point in the fiber of $\psi_{\mathcal{L}}$ is a rational function of~$\mathbf{M}$.

\begin{figure}[htbp]
    \centering
    \includegraphics[width=0.45\textwidth]{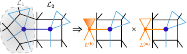}
    \caption{LS discriminant recursion of the rational  
    pentabox.}
    \label{fig:rat_pentabox}
\end{figure}

Physically, when two external lines intersect, the corresponding external momentum at that vertex of the dual graph is on shell; otherwise it is off shell, while the propagators remain massless; see Fig.~\ref{fig:rat_pentabox}. Thus one starts from fully off-shell kinematics and imposes on-shell conditions until $\mathcal{L}$ is rational. In \cite[Section 8]{Our_Landa_on_Gr} we study rationality for special graphs $G$.

\smallskip

\textit{Example ($\ell=1$)} 
The $4$-mass box is made rational by setting $\langle M_1M_2 \rangle =0$. This is the \textit{three-mass box}: $G=K_1$, $u=(4)$ and $H_u$ being one edge. 
The two transversals to $M_1=(12)$, $M_2=(23)$, $M_1=(45)$ and $M_1=(67)$ are
$
L^{({\bf b})}= (245)\star (267) \, , \
L^{({\bf w})}=\bigl((123)\star (45),(123)\star (67)\bigr) $,
and $\Delta_{K_1,(4)}$ becomes $\Delta_{\mathcal{L}} = \langle L^{(\mathbf{b})} L^{(\mathbf{w})} \rangle = \langle 245|13|672\rangle^2 $. This is
the square of a cluster variable in $\Gr(4,7)$. 

Cluster algebras famously govern the symbol alphabet of planar $\mathcal{N}=4$ SYM amplitudes. We formulate this phenomenon in the present setting as follows.

\medskip
\noindent\textbf{Cluster Factorization Conjecture.}
Let $\mathcal{L}$ be a planar leading rational Landau diagram. Then $\Delta_\mathcal{L}$ factors into a product of cluster variables for $\Gr(4,n)$\vspace{1.5mm}.

The origin of cluster algebras in this context was until now mysterious. 
We suggest an explanation:
\emph{cluster variables arise as discriminants encoding collisions of rational leading singularities}.
Moreover,~\eqref{eq:recursion} reveals a recursive mechanism behind cluster factorization: if $\Delta_{\mathcal{L}_1}$ and ${\rm LS}_{\mathcal{L}_2}$ are products of cluster variables, and each substitution map $\varphi_r^*$ sends cluster variables to cluster variables---that is, if each $\varphi_r^*$ is a \emph{cluster quasi-homomorphism}---then ${\rm LS}_{\mathcal{L}}$ factors into cluster variables. We now show this mechanism for trees $G$ using the three-mass box as seed.

Let $\mathcal{L}$ be the rational pentabox of Fig.~\ref{fig:rat_pentabox}: $G=K_2$, $u=(4,3)$ and $H_u$ with two edges. Vertex $1$ has the external lines
$a_1a_2,\, a_2a_3,\, a_4a_5,\, a_6a_7$,
and vertex $2$ has the external lines
$b_1b_2,\, b_2b_3,\, b_4b_5$.
These are ordered.
By the recursion, the LS discriminant $\Delta_{\mathcal{L}}$ factors as follows:
\begin{equation}\label{eq:pentabox_rat}
\begin{small}
\begin{aligned}
    &\langle b_2 b_4 b_5 | b_1 b_3 | b_2 , (a_4 a_1 a_2) \star (a_4 a_6 a_7) \rangle^2   \\
    \cdot   &\langle b_2 b_4 b_5 | b_1 b_3 | b_2 , (a_3 a_4 a_5) \star (a_1 a_2) , (a_3 a_4 a_5) \star (a_6 a_7) \rangle^2 \, .
\end{aligned}
\end{small}
\end{equation}
Both factors in~\eqref{eq:pentabox_rat} are cluster variables for $\Gr(4,12)$. 

More generally, let $\mathcal{L}=\mathcal{L}_1\cup\mathcal{L}_2$ be a Landau diagram reducible by the three-mass box $\mathcal{L}_1$. By~\cite[Prop 12.3]{Our_Landa_on_Gr},
\begin{equation}\label{eq:3mb_rec}
\Delta_\mathcal{L}
=
\varphi_{\mathbf{b}}^*\Delta_{\mathcal{L}_2}\cdot
\varphi_{\mathbf{w}}^*\Delta_{\mathcal{L}_2} \, .
\end{equation}
The $\varphi_\sigma^*$ in~\eqref{eq:3mb_rec} are cluster quasi-homomorphisms~\cite[Lemma 12.5]{Our_Landa_on_Gr}. For example, $\varphi_{\mathbf{b}}^*$ is closely related to \emph{BCFW promotion}~\cite[Section 4]{m4tiling}. This fact yields an inductive proof that~\eqref{eq:pentabox_rat} is a product of cluster variables. 

\smallskip
\noindent\textbf{Theorem} (\cite[Thm 12.6]{Our_Landa_on_Gr}).
The Cluster Factorization Conjecture is true if 
the underlying graph $G$ is a tree\vspace{1.5mm}.

The first Landau diagram whose graph $G$ is not a tree is
$\mathcal{L}=G_u\cup H_u$ with $G=K_3$, $u=(3,3,3)$, and one pendant triangle per vertex. We restrict to the irreducible component of $V_G$ where the three lines are concurrent. The Landau map has $8$ rational points in its fiber, one for each coloring (black {\bf b} for concurrent or white {\bf w}
for coplanar) of the three pendant triangles.
The LS discriminant $\Delta_{\mathcal{L}}$ factors into $12$ irreducible factors, one for each edge of the $3$-cube labeled by
$\{{\bf b},{\bf w}\}^3$.
Each factor is the square of a cluster variable. Let
$a_1a_2,\ a_2a_3,\ a_4a_5$
denote the external lines at vertex $1$, and similarly with $b_i$ and $c_i$ at vertices $2$ and $3$. The $15$ vectors defining these lines form a totally positive $4\times15$ matrix ${\bf z}$, whose columns are ordered as
$a_1,a_2,\ldots,a_5,\ b_1,b_2,\ldots,b_5,\ c_1,c_2,\ldots,c_5$.
One of the factors is $\langle a_2 a_4 a_5 | a_1 a_3 | a_2,\,(b_2 b_4 b_5)\star(c_2 c_4 c_5)\rangle^2$,
which is a cluster variable for $\Gr(4,15)$. Each factor can be written as the pairing $\langle L_i^{(\sigma_1)}L_i^{(\sigma_2)}\rangle$, where the colorings $\sigma_1$ and $\sigma_2$ differ in exactly one position $i$.  Similarly, the coplanar component has $12$ analogous factors. 

This suggests the following general picture. For planar rational Landau diagrams, the rational points in the fibers correspond to colorings of the $\ell$ pendant triangles, according to whether they form triples of coplanar or concurrent lines. The LS discriminant is expected to factor into cluster variables, each associated with a pair of colorings $\sigma_1,\sigma_2$ that differ at a single position $i$, see~\cite[Conj 8.12]{Our_Landa_on_Gr}. The corresponding factor is given by the pairing $\langle L_i^{(\sigma_1)}L_i^{(\sigma_2)}\rangle$, for suitable representatives of $L_i^{(\sigma_1)},L_i^{(\sigma_2)}$.

The mechanism underlying both positivity and cluster structure of LS discriminants is the recursion \eqref{eq:recursion}, together with the 
interplay with the substitution maps. By relating line incidence varieties to positroid varieties and amplituhedron maps, we show in \cite[Sections 10-12]{Our_Landa_on_Gr} that these substitution maps coincide with \emph{promotion maps} \cite{PlabicTangles} between Grassmannians defined from plabic graphs.
General promotion maps are conjectured to preserve positivity and, in the rational regime, to define cluster quasi-homomorphisms, referred to as \emph{cluster promotion maps} \cite[Conjectures 4.16, 10.13]{PlabicTangles}. This leads~to:

\smallskip

\noindent\textbf{Substitution Maps Conjecture.}
Let $\mathcal{L}$ be a planar Landau diagram reducible by $\mathcal{L}_1$, as in \eqref{eq:recursion}. Then the substitution maps $\varphi_r$ are copositive, i.e.\ they map positive lines to positive lines. Moreover, in the rational regime, the pullbacks $\varphi_r^*$ are cluster quasi-homomorphisms.

\vspace{-0.5em}
\section{Outlook}
\vspace{-1em}

This Letter, together with the papers~\cite{Our_lines_1,Our_Landa_on_Gr}, develops a rigorous framework for Landau analysis on the Grassmannian. We uncover a recursive mechanism governing a large class of Landau singularities, bridging their algebro-geometric description as discriminants and resultants with positivity and cluster structures seen in planar $\mathcal{N}=4$ SYM. 
Directions for future research are outlined in~\cite[Section 13]{Our_Landa_on_Gr}. These include the extension of our framework to subleading singularities, the relation between the Landau analysis in the Grassmannian and the loop amplituhedron, and  cluster adjacencies from discriminants across different strata of Landau singularities.

\bigskip 

\noindent {\bf Acknowledgement}:~We were supported by
the European Research Council synergy grant UNIVERSE+, \begin{small}101118787.\end{small}
$\!\!$ \begin{scriptsize}Views~and~opinions expressed
are however those of the~authors only and do not necessarily reflect those of the European Union or the
European
Research Council Executive Agency. Neither the European Union nor the granting authority
can be held responsible for them.
\end{scriptsize}

\bibliographystyle{apsrev4-2}

\begin{thebibliography}{99}


\bibitem{The_amplituhedron}
N.~Arkani-Hamed and J.~Trnka,
\newblock ``The Amplituhedron,''
\newblock {\em Journal of High Energy Physics}, vol.~2014, no.~10, article 030, October 2014,

\bibitem{Higher_roots}
J.~L.~Bourjaily, C.~Vergu, and M.~von Hippel,
\newblock ``Landau singularities and higher-order polynomial roots,''
\newblock {\em Phys. Rev. D} {\bf 108}  (2023) article 085021.

\bibitem{BT} 
P.~Breiding and S.~Timme:
{\em HomotopyContinuation.jl: A package for homotopy continuation in Julia},
Mathematical Software -- ICMS 2018, 458--465, Springer, 2018.

\bibitem{Cachazo:2008vp}
F.~Cachazo,
\newblock Sharpening the Leading Singularity,
\newblock {\tt arXiv:0803.1988}.

\bibitem{Caron-Huot:2025recursive}
S.~Caron-Huot, M.~Correia and M.~Giroux,
\newblock Recursive Landau Analysis,
\newblock Phys. Rev. Lett. {\bf 135} (2025) 131603.

\bibitem{CHDDDFGvHMP}
S.~Caron-Huot, L.~J.~Dixon, F.~Dulat, M.~von Hippel, A.~J.~McLeod and G.~Papathanasiou,
\newblock The six-gluon amplitude in planar $\mathcal{N}=4$ super-Yang--Mills theory at six and seven loops,
\newblock JHEP {\bf 08} (2019) 016.

\bibitem{Correia:2025sofia}
M.~Correia, M.~Giroux and S.~Mizera,
\newblock SOFIA: Singularities of Feynman Integrals Automatized,
\newblock {\tt arXiv:2503.16601}.

\bibitem{Dennen:2017}
T.~Dennen, I.~Prlina, M.~Spradlin, S.~Stanojevic and A.~Volovich,
\emph{Landau singularities from the amplituhedron},
JHEP \textbf{06} (2017) 152.

\bibitem{Drummond:2010qh}
J.~M.~Drummond, J.~M.~Henn, G.~P.~Korchemsky and E.~Sokatchev,
\emph{Simple loop integrals and amplitudes in $\mathcal{N}=4$ SYM},
JHEP \textbf{01} (2011) 064.

\bibitem{DFG}
J.~Drummond, J.~Foster and \"O.~G\"urdo\u{g}an,
\newblock Cluster adjacency properties of scattering amplitudes,
\newblock Phys.\ Rev.\ Lett.\ {\bf 120} (2018) 161601.

\bibitem{Eden:1966dn}
R.~J.~Eden, P.~V.~Landshoff, D.~I.~Olive and J.~C.~Polkinghorne,
\newblock The Analytic S-Matrix,
\newblock Cambridge University Press (1966).


\bibitem{EZLPTSW}
C.~Even-Zohar, T.~Lakrec, M.~Parisi, M.~Sherman-Bennett,
R.~Tessler and L.~Williams:
{\em BCFW tilings and cluster adjacency for the amplituhedron},
Proc.~Natl.~Acad.~Sci.~U.S.A. {\bf 122} (2025) e2408572122.

\bibitem{m4tiling}
C.~Even-Zohar, T.~Lakrec, M.~Parisi, M.~Sherman-Bennett, R.~Tessler and L.~Williams:
{\em Cluster algebras and tilings for the $m=4$ amplituhedron},
{\tt arXiv:2310.17727}.

\bibitem{PlabicTangles}
C.~Even-Zohar, M.~Parisi, M.~Sherman-Bennett, R.~Tessler and L.~Williams,
\emph{Plabic Tangles and Cluster Promotion Maps},
J. Algebra \textbf{695} (2026).

\bibitem{Fevola:2024principal}
C.~Fevola, S.~Mizera and S.~Telen,
\newblock Principal Landau determinants,
\newblock Comput.\ Phys.\ Commun.\ {\bf 303} (2024) 109278.

\bibitem{Fevola:2024revisited}
C.~Fevola, S.~Mizera and S.~Telen,
\newblock Landau singularities revisited: Computational algebraic geometry for Feynman integrals,
\newblock Phys.\ Rev.\ Lett.\ {\bf 132} (2024) 101601.

\bibitem{Fraser} 
C.~Fraser,
{\em Quasi-homomorphisms of cluster algebras},
Adv.~Appl.~Math.~{\bf 81} (2016) 40--77.

\bibitem{Golden:2013xva}
J.~Golden, A.~B.~Goncharov, M.~Spradlin, C.~Vergu and A.~Volovich,
\newblock Motivic amplitudes and cluster coordinates,
\newblock JHEP {\bf 01} (2014) 091.

\bibitem{GP23}
\"O.~G\"urdo\u{g}an and M.~Parisi:
{\em Cluster patterns in Landau and leading singularities via the amplituhedron},
Ann.~Inst.~Henri Poincar\'e D {\bf 10} (2023) 299--336.

\bibitem{Sequential_Discontinuitites}
H.~S.~Hannesdottir, A.~J.~McLeod, M.~D.~Schwartz, and C.~Vergu,
\newblock ``Constraints on sequential discontinuities from the geometry of on-shell spaces,''
\newblock {\em JHEP}, vol.~07, article 236, 2023.

\bibitem{Hannesdottir:2022iterated}
H.~S.~Hannesdottir, A.~J.~McLeod, M.~D.~Schwartz and C.~Vergu,
\newblock Implications of the Landau equations for iterated integrals,
\newblock Phys.\ Rev.\ D {\bf 105} (2022) L061701.

\bibitem{Hannesdottir:2025bootstrap}
H.~S.~Hannesdottir, A.~J.~McLeod, M.~D.~Schwartz and C.~Vergu,
\newblock Applications of the Landau bootstrap,
\newblock Phys.\ Rev.\ D {\bf 111} (2025) 085003.

\bibitem{He:2024schubert}
S.~He, X.~Jiang, J.~Liu and Q.~Yang,
\newblock Landau-based Schubert analysis,
{\tt arXiv:2410.11423}.

\bibitem{helmer2024landau}
M.~Helmer, G.~Papathanasiou and F.~Tellander:
{\em Landau singularities from Whitney stratifications},
{\tt arXiv:2402.14787}. 

\bibitem{hodges2013eliminating}
A.~Hodges,
\newblock ``Eliminating spurious poles from gauge-theoretic amplitudes,''
\newblock {\em Journal of High Energy Physics}, vol. 2013, no. 5, article 135, 2013.

\bibitem{Our_lines_1}
B.~Hollering, E.~Mazzucchelli, M.~Parisi, and B.~Sturmfels,
\newblock ``Varieties of Lines in 3-Space,''
{\tt arXiv:2511.21333}.

\bibitem{Our_Landa_on_Gr}
B.~Hollering, E.~Mazzucchelli, M.~Parisi, and B.~Sturmfels,
\newblock ``Landau Analysis on the Grassmannian,''
\newblock to appear.

\bibitem{Landau1960}
L.~D.~Landau,
\newblock ``On analytic properties of vertex parts in quantum field theory,''
\newblock {\em Nuclear Physics}, vol.~13, pp.~181--192, 1960.

\bibitem{Mizera:2022landau}
S.~Mizera and S.~Telen,
\newblock Landau discriminants,
\newblock JHEP {\bf 08} (2022) 200.

\bibitem{Pham2011}
F.~Pham,
\newblock {\em Singularities of Integrals: Homology, Hyperfunctions and Microlocal Analysis},
\newblock Springer, 2011.

\bibitem{Ranestad:2024adjoints}
K.~Ranestad, R.~Sinn and S.~Telen,
\newblock Adjoints and canonical forms of tree amplituhedra,
\newblock Math. Scand.\ {\bf 130} (2024).

\bibitem{SS}
J.~Sidman and A.~Lee--St.John:
\emph{Frameworks in Motion -- An Introduction to Rigidity},
Manuscript for an undergraduate textbook,
  \url{https://rigidity-theory.web.app}.

\bibitem{Weinzierl2022}
S.~Weinzierl: {\em Feynman Integrals}, Springer Verlag, 2022.

\end{thebibliography}

\end{document}